\documentclass[12pt,preprint]{aastex}
\usepackage{epsfig}
\usepackage{multirow}
\usepackage{longtable}
\usepackage{natbib}
\usepackage{rotating}
\usepackage{amsmath, amsthm, amssymb, amsfonts}

\newcommand{\agile}{\textit{Agile}}
\newcommand{\rband}{$r'$--band}

\newcommand{\mearth}{M$_\oplus$}

\newcommand{\tmcmc}{\texttt{TMCMC}}
\newcommand{\mtqlong}{\texttt{MultiTransitQuick}}
\newcommand{\mtq}{\texttt{MTQ}}

\newcommand{\boldtheta}{\boldsymbol{\theta}}

\newcommand{\gaussian}{\mathcal{G}}
\newcommand{\redchi}{$\chi^{2}/\nu$}

\newcommand{\eq}{Eq.}
\newcommand{\ie}{\textit{i.e.}}
\newcommand{\apostle}{APOSTLE}
\newcommand{\hst}{\textit{HST}}
\newcommand{\corot}{\textit{CoRoT}}

\newcommand{\keptel}{\textit{Kepler}}

\newcommand{\asec}{$"$}
\newcommand{\degree}{$^{\circ}$}

\newcommand{\tres}{TrES-3}

\newcommand{\tresb}{TrES-3b}
\newcommand{\gjb}{GJ 1214b}

\newcommand{\ldc}{LDC}
\newcommand{\sect}{$\S$}

\newcounter{parnum}
\newcommand{\N}{%
   \noindent\refstepcounter{parnum}%
    \makebox[\parindent][l]{\textbf{\Roman{parnum}.}}}
\begin{document}
\title{APOSTLE: Eleven Transit Observations of \tresb}
\author{
  P.~Kundurthy\altaffilmark{1},
  A.C.~Becker\altaffilmark{1},
  E.~Agol\altaffilmark{1},
  R.~Barnes\altaffilmark{1,2},
  B.~Williams\altaffilmark{1}
}
\altaffiltext{1}{Astronomy Department, University of Washington, Seattle, WA 98195}
\altaffiltext{2}{Virtual Planetary Laboratory, USA}

\begin{abstract}
The Apache Point Survey of Transit Lightcurves of Exoplanets (APOSTLE) observed eleven transits of \tresb\ over two years in order to constrain system parameters and look for transit timing and depth variations. We describe an updated analysis protocol for \apostle\ data, including the reduction pipeline, transit model and Markov Chain Monte Carlo analyzer. Our estimates of the system parameters for \tresb\ are consistent with previous estimates to within the 2$\sigma$ confidence level. We improved the errors (by 10--30\%) on system parameters like the orbital inclination ($i_{\text{orb}}$), impact parameter ($b$) and stellar density ($\rho_{\star}$) compared to previous measurements. The near-grazing nature of the system, and incomplete sampling of some transits, limited our ability to place reliable uncertainties on individual transit depths and hence we do not report strong evidence for variability. Our analysis of the transit timing data show no evidence for transit timing variations and our timing measurements are able to rule out Super-Earth and Gas Giant companions in low order mean motion resonance with \tresb.
\end{abstract}
\keywords{eclipses, stars: planetary systems, planets and satellites: fundamental parameters,individual: \tresb}

\section{Introduction} \label{sec_intro}
When an extrasolar planet eclipses its host star, the event is referred to as a transit or primary eclipse. During a transit, observers can detect dips in starlight caused by the planet obscuring a portion of the stellar disk when it passes in front of the star. The transit method applies to those systems where the orbital inclination of a planet is close to $90$\degree\ (\ie\ edge-on) with respect to the observer's sky-plane. The first transit observations were made by \citet{charbonneau00}. As of August 2012, more than 200 planets (exoplanet.eu) have been detected using the transit method. The search for new exoplanetary systems via transits has advanced to space-based missions with the launch of the European Space Agency's (ESA) \corot\ satellite \citep{fridlund06} and NASA's \keptel\ mission \citep{borucki10}. The objective of transit detection and follow-up is mainly to catalog and improve measurements of system parameters. Studying the characteristics of extrasolar planetary systems is crucial for developing theories of planet formation that can adequately explain the origin and evolution of all planetary systems (including our own).

The target discussed in this paper, \tresb\, is a Hot-Jupiter with one of the shortest orbital periods known \citep[P=1.3 days][]{odonovan07} among the exoplanets. The planet orbits a G-type star (T$_{eff}$ = 5720K), and has a mass and radius of $M_{p} =$1.92 $M_{\text{Jup}}$ and $R_{p} = $1.29 $R_{\text{Jup}}$, respectively \citep{odonovan07}. Due to its large impact parameter, \tresb's transit has more of a v-shape than the typical u-shape. However, the transit is not grazing and the ability of various observers to consistently measure the transit parameters seem to indicate that the planet's disk completely enters the stellar disk during the transit \citep{odonovan07,sozzetti09, gibson09}. A blended eclipsing binary, which could account for the v-shape, has been ruled out from the radial velocity analysis \citep{odonovan07}. However, the v-shape of the transit makes measurements of transit properties challenging at shorter wavelengths, where stellar limb-darkening further degrades the trapezoidal u-shape of the transit. The level of insolation received on the dayside of \tresb\ due to its proximity to its host star should place it in the class of warm Hot-Jupiters with a temperature inversion in its atmosphere \citep{hubeny03,burrows07,fortney08}. However, secondary eclipse observations from the ground and space have shown that the evidence for an inversion layer is not strong \citep{demooijsnellen09,fressin10}. \tresb's lack of an inversion layer is consistent with the idea that UV radiation from chromospherically active stars (like \tresb's parent star) systematically destroys absorbers in the upper planetary atmosphere, thereby preventing the formation of an inversion layer \citep{knutson10}.

The ultra close-in orbit of \tresb\ also makes it a great target for follow-up transit monitoring. It has been suggested that planets with very short periods are likely to be falling into their host stars \citep{jackson09}. Some have looked for transit timing variations indicative of such orbital decay \citep[for e.g. OGLE-TR-56b][]{adams11}. \tresb\ is similar to OGLE-TR-56b in many regards, including the orbital period ($< 1.5$ days), size ($\sim 1 R_{\text{Jup}}$) and impact parameter ($b > 0.8$). Though there has been no strong evidence for TTVs for OGLE-TR-56b, it is one object that warrants long term study \citep{adams11}. Transit monitoring of \tresb\ by other teams has so far confirmed its linear orbital ephemeris \citep{sozzetti09,gibson09}. A search for additional transiting companions using the EPOXI mission has yielded no detections, since the likelihood of detecting a planetary sibling in a coplanar orbit with \tresb\ is lower for an outer planet given the inclination of the system \citep{ballard11}; inner planets are not expected since \tresb\ is already quite close to its parent star.

In this paper we report observations of eleven transits of \tresb, taken as part of the Apache Point Observatory Survey of Transit Lightcurves of Exoplanets (APOSTLE). The \apostle\ program is a follow-up transit monitoring program designed to obtain high-precision relative photometry on known transiting systems in order to refine measurements of system parameters and transit times \citep[see e.g. \gjb][]{kundurthy11}. In \sect\ \ref{sec_obs} we outline our observations, and in \sect\ \ref{sec_pipeline} we describe the data reduction. The two sections which follow, \sect\ \ref{sec_mtq} and \sect\ \ref{sec_tmcmc}, outline the transit model and the Markov Chain Monte Carlo (MCMC) analyzer respectively. In \sect\ \ref{sec_sysparams} we present our estimates of the system parameters for \tresb\ and in the subsections \sect\ \ref{sec_tdv} and \sect\ \ref{sec_ttv} we present results from our study of transit depth variations (TDVs) and transit timing variations (TTVs). Finally, in \sect\ \ref{sec_conclusions} we summarize our findings.

\section{Observations} \label{sec_obs}
\tres\ was observed by APOSTLE over a time span of 2 years between the summer of 2009 and the fall of 2011. All observations of \tres\ were carried out using \agile\, a high-speed frame-transfer photometer \citep{mukadam12}, on the ARC\footnote{Astrophysical Research Consortium} 3.5m telescope at Apache Point, New Mexico. The \agile\ CCD has no dead time, as the charge is transferred to an adjoining array for read-out. All observations for \tres\ were made using \agile's medium-gain, slow read-out (100kHZ) mode, with the charge read-out at 45sec intervals. The filter used for the observations was the \rband\, similar to the SDSS\footnote{Sloan Digital Sky-Survey} $r$ filter \citep[with central wavelength, $\lambda_{0} = $ 626nm,][]{fukugita96}. This observing filter is bluer than typical filters used for transit observations since \agile\ is a blue sensitive CCD that is affected by a strong fringe pattern at longer wavelengths \citep{mukadam12}. The summary of the 11 \rband\ observations is given in Table \ref{table_ObsSum_TRES3}. Observations were made by adjusting the focus on the secondary mirror to smear the stellar Point Spread Functions (PSFs) across multiple pixels, which minimizes the systematics caused by pixel-to-pixel wandering of the PSF. The long read-out (exposure time) also allowed for a greater count rate that maximized the signal-to-noise per image. The count rate was kept below Agile's non-linearity limit of $\sim$52k ADU and well below its saturation level of 61k ADU by small adjustments to the telescope's secondary focus.

\begin{sidewaystable}[!h]
\begin{center}
\caption{\label{table_ObsSum_TRES3} APOSTLE Observing Summary for TRES3}
\begin{tabular}{ccrccccrrcc}
\hline \hline
T\# & UTD & Obs. Cond. & Filter & Exp. & Phot. Ap. &RMS (ppm) & \%Rej. & Flux Norm. & Error Scaling \\
(1) & (2) & (3) & (4) & (5) & (6) &(7) & (8) & (9) & (10) \\
\hline
1&2009-05-14&Clear&r'&45&30&777&2\%&1.9179&0.5618\\
2&2009-06-13&Poor Weather&r'&45&22&610&13\%&1.9083&0.5135\\
3&2010-03-22&Clear&r'&45&20&962&$ < 1\%$&1.7788&0.7436\\
4&2010-05-16&Clear&r'&45&21&877&1\%&1.8954&0.7769\\
5&2010-06-02&Poor Weather&r'&45&30&703&$ < 1\%$&1.8670&0.2287\\
6&2010-10-12&Clear&r'&45&24&896&$ < 1\%$&1.8832&0.7804\\
7&2011-03-24&Clear&r'&45&19&644&1\%&1.9809&0.5644\\
8&2011-04-27&Poor Weather&r'&45&14&1022&1\%&1.8932&0.9787\\
9&2011-05-14&Poor Weather&r'&45&15&2960&$ < 1\%$&1.8887&2.8965\\
10&2011-06-21&Clear&r'&45&26&1221&$ < 1\%$&1.9227&0.9683\\
11&2011-08-24&Clear&r'&45&19&1240&7\%&1.8016&0.7597\\
\hline
\end{tabular}
 \footnotesize \begin{tabular}{l}
(1) Transit Number, (2) Universal Time Date, (3) Observing Conditions, (4) Observing Filter, (5) Exposure Time (seconds) \\
(6) Optimal Aperture Radius (pixels), (7) Scatter in the residuals \\
(8) \% frames rejected due to saturation or other effects, (9) Flux normalization between the target and comparison star \\
(10) The factor by which the photometric errors were scaled \\
\end{tabular}
 \normalsize \end{center}
\end{sidewaystable}

The parent star of \tresb, GSC 03089-00929 (\tres) is a G type star with a Johnson R magnitude of 12.2. The comparison star used for the relative photometry was USNO-B1.0 1275-0332540 situated $\sim$ 90\asec\ away, with Johnson R of 12.9 \citep{monet03}, which was the next brightest star in \agile's FOV. Several studies that use detectors with larger FOVs than \agile\ typically use many (N $>>$ 1) comparison stars. For relative aperture photometry, lightcurve precision is limited by the signal-to-noise achieved from aperture extraction on the faintest star in the set. The small FOV of \agile\ meant that most other comparison stars in the field were too faint to provide the adequate signal-to-noise on the final lightcurve. PSF photometry is a solution to the problem of using faint comparison stars since the stellar and background components can be constrained accurately. However due to the difficulty in modeling the complex defocused PSF of \apostle\ observations we did not opt for this route. We used a circular aperture for photometry (see details in \sect\ \ref{sec_pipeline}) and hence the target and brightest companion produced the best results. Our uncalibrated differentrial photometry shows that \tres\ was the brighter of the two by a factor of $\sim$ 1.8 in the \rband. We also note variability in the uncalibrated flux between \tres\ and its reference star, with the maximum difference being $\sim$20\% between the highest (\#7 UTD 2011-03-24) and lowest (\#3 UTD 2010-03-22) values. The observations were made over a variety of observing conditions (Column `Obs. Conditions' in Table \ref{table_ObsSum_TRES3}). The listed nights include both complete and partial transits (where data were lost due to poor weather, or issues with the instrument). Some of these partial transits include UTD 2009-05-14 ($\# 1$), 2009-06-13 ($\# 2$) and 2010-06-02 ($\# 10$). Small portions of the in-eclipse data were lost for the transits on UTD 2010-03-22 ($\# 3$) and UTD 2011-04-27 ($\# 8$), and the night of UTD 2011-05-14 ($\# 9$) had exceptionally poor observing conditions (seeing $> 2$\asec). Even though these data will affect the fit, we include them in the analysis since they can be used to determine transit times and other system parameters.

\section{APOSTLE Pipeline} \label{sec_pipeline}
The \apostle\ project used a customized data reduction pipeline, written in the Interactive Data Language (IDL) to process data from \agile. The pipeline performs standard image processing steps like dark subtraction and flat fielding, but also implements non-linearity corrections unique to \agile. The pipeline also creates an uncertainty map of the processed images by propagating pixel-to-pixel errors through each step of the reduction. In addition to the photon counting errors and read-noise from the raw images, the pipeline propagates the variance on the master dark and master flat during the reduction. Errors were also propagated for those pixels where the counts exceeded the non-linearity threshold ($\sim$52k counts) using the uncertainties in an empirically derived non-linearity correction function. Frames where pixels inside a photometric aperture exceeded \agile's saturation limit of 61k were rejected. Images at the other extreme, where the stars were obscured by clouds, and resulted in low signal to noise measurements were also rejected (\ie\ where photometric errors were $> 5000$ppm). The fraction of rejected frames per night is listed in Column (9) `$\%$Rej.' in Table \ref{table_ObsSum_TRES3}.

We used SExtractor \citep{bertinarnouts96} to derive initial centroids of our defocused stars. This software allows the use of a customized PSF kernel, so we used a `donut'-shaped detection kernel for our defocused data. Coordinates obtained from SExtractor were then used for circular aperture photometry with the PHOT task in IRAF's NOAO.DIGIPHOT.APPHOT package. We derived flux estimates from a range of circular apertures with radii between 5--50 pixels, at intervals of 1 pixel, by simply summing the counts in these apertures. An outlier--rejected global median on the frame was used as the sky estimate, which is removed to derive the instrumental flux of the stars. To derive photometric errors, we extracted counts from the error frames using the same centroids and apertures used for photometry on the target frames. The lightcurves are generated by dividing the instrumental flux of the target star by the comparison star and then dividing the entire lightcurve by the median out-of-eclipse flux level. At this stage there may still be systematic trends in the lightcurve which have not been removed by reduction; for example, differential extinction due to airmass variation or photometric variation due to centroids wandering over pixels of varying sensitivities (e.g. due to small imperfections in the flat-fielding). Thus, for each image we also extracted a set of nuisance parameters which are used to compute a correction function (\ie\ detrending function). For the \tres\ data we found that, (i) the airmass, (ii) the global median sky, (iii) the centroid positions of the target star, and the sum of counts in the photometric aperture for (iv) the master dark and (v) the master flat, showed trends that corresponded to trends seen in the lightcurves. The airmass was derived from the image headers, while the global sky and the centroid positions are derived from the photometry on the science frames. The sum of counts in the master dark and master flat are derived from photometry on the master dark and flat using the centroids and apertures used for photometry on the science frames. The correction function ($F_{cor}$) is modeled as a linear sum of nuisance parameters as described by the following equation:
\begin{equation}
F_{cor,i} = \sum_{k=1}^{N_{\text{nus}}} c_{k} X_{k,i},
\label{eq_detrending}
\end{equation}
where $X_{k,i}$ are the nuisance parameters, $c_{k}$ are the corresponding coefficients. The index $k$ counts over the number of nuisance parameters $N_{\text{nus}}$, and the index $i$ denotes the transit number. The detrending coefficients are chosen by minimizing the $\chi^2$ between the observed data ($O$), a model function ($M$) and correction function,
\begin{equation}
\chi^2 = \sum_{j}^{N_{\text{all}}} \frac{(O_{j} - M_{j} - F_{\text{cor},j})^2}{\sigma_{j}^2}
\label{eq_chisq_multi}
\end{equation}
here $j$ is the index which counts over the total number of data points ($N_{\text{all}}$), when all transits are stacked. During photometry we use a set of trial model parameters (based on values from the literature) to remove the transit lightcurve from the data. We use a linear least-squares minimizer to fit for the coefficients of $F_{cor,i}$ (Eq. \ref{eq_detrending}), which removes any correlated trends that remain in the model-subtracted lightcurve. The resulting residuals are used for selecting the optimal aperture from which to extract the photometry. The aperture where the scatter in the residuals is minimized was chosen as the optimal aperture. Smaller apertures do not completely sample the flux from the stars, and at larger apertures one loses signal-to-noise due to the fact that one accumulates fewer counts from the star, compared to sky counts. The optimal aperture typically fell between 5--50 pixel radii. We present the complete list of optimal apertures in Column (6) in Table \ref{table_ObsSum_TRES3}. The sizes of the typical apertures were always a few times larger than the typical half-width at half maximum of the stellar PSFs. The photometric precision, which is simply the scatter in the residuals, is also listed in Table \ref{table_ObsSum_TRES3} in Column (7).

\begin{table}
\begin{center}
\caption{\label{table_LCdata_TRES3} APOSTLE Lightcurve data* for TrES-3}
\begin{tabular}{ccccc}
\hline \hline 
T\# & T-T0 & Norm. Fl. Ratio & Err. Norm. Fl. Ratio & Model Data \\
(1) & (2) & (3) & (4) & (5) \\
 \hline
1&-0.0490181&0.999407424&0.000934904&1.000000000\\
1&-0.0479764&1.000889277&0.000936546&1.000000000\\
1&-0.0474555&1.001329344&0.000936130&1.000000000\\
1&-0.0469347&1.002419716&0.000941452&1.000000000\\
1&-0.0455226&1.000658752&0.000936362&1.000000000\\
1&-0.0444809&0.998944209&0.000923458&1.000000000\\
1&-0.0434392&0.998014984&0.000914333&1.000000000\\
1&-0.0429184&0.996495746&0.000917120&1.000000000\\
1&-0.0423975&0.999418076&0.000914379&1.000000000\\
1&-0.0418767&1.000283709&0.000908789&1.000000000\\
 . & . & . & . & . \\
 . & . & . & . & . \\
\hline
\end{tabular}
\footnotesize \begin{tabular}{l} 
*The data are presented in their entirety as an online-only table. \\
(1) Transit Number, (2) Time Stamps - Mid Transit Times (BJD) \\
(3) Normalized Flux Ratio, (4) Error on Normalized Flux Ratio \\
(5) Model Data \\
 \end{tabular}
\end{center}
\end{table}

\clearpage
\begin{figure}[!h] 
\begin{center} 
\includegraphics[width=0.75\textwidth]{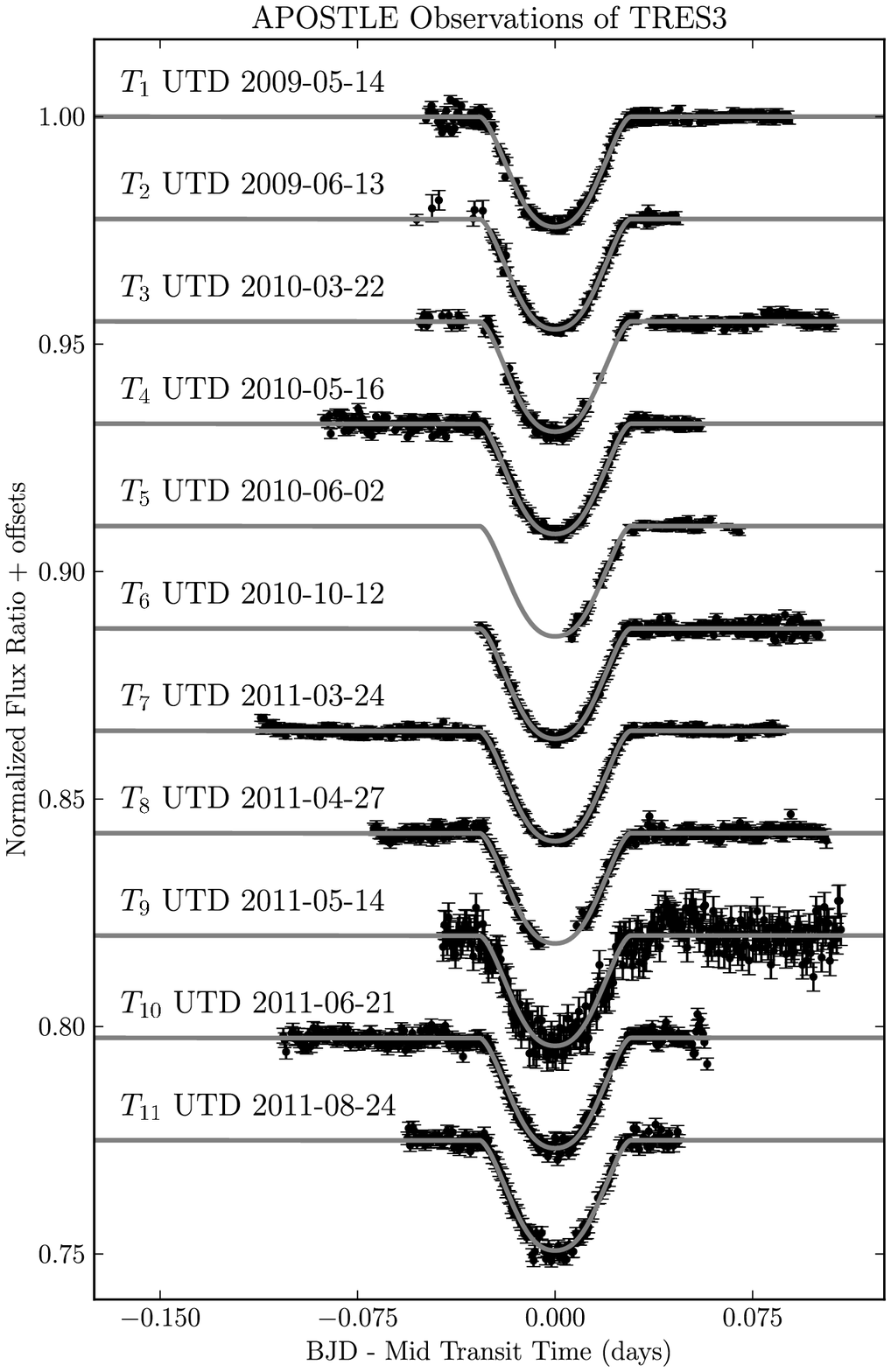}
\caption{\label{figure_LC_TRES3}Eleven \rband\ lightcurves of \tresb. The vertical axis is in normalized flux ratio units. The horizontal axis shows time from the mid-transit time in days, computed by subtracting the appropriate mid-transit time for each transit from the best-fit values in the Fixed LDC chain.}
\end{center} 
\end{figure}

The 11 transits of \tresb\ are shown in Figure \ref{figure_LC_TRES3} in normalized flux ratios (with offsets for clarity). The plotted data result from the data reduction and model fitting processes described in sections \sect\ \ref{sec_pipeline}, \sect\ \ref{sec_mtq} and \sect\ \ref{sec_tmcmc} respectively. The plotted data are presented as an online-only table.

\section{Multi Transit Quick} \label{sec_mtq}
We developed a transit model called \mtqlong\ (\mtq) in PYTHON, which is based on the analytic lightcurve models presented in \citet{mandelagol02}, and the PYTHON implementation of some of its functions \citep[from EXOFAST by][]{eastman12}. The set of transit parameters used by \mtq\ include the transit duration ($t_{T}$), the limb-crossing duration ($t_{G}$) and the times of mid-transit ($T0$). The parameters $t_{T}$ and $t_G$ are the same as $T$ and $\tau$ from \citet{carter08}. The shape of the transit is characterized by the transit depth ($D$) and the stellar limb-darkening parameters $v_1$ and $v_2$. The parameter $D$ simply represents the maximum depth of the transit trough at conjunction, which depends on the ratio of the area of the disks of the planet to the star ($R^2_{p}/R^2_{\star}$) and the ratio of the disk-averaged stellar intensity at conjunction $I(b)$ to the un-obscured disk-averaged intensity at conjunction for an impact parameter of zero, $I(0)$. The transit depth $D$ is:
\begin{subequations}
\begin{equation}
D = \frac{R^2_{p}}{R^2_{\star}} \frac{I(b)}{I(0)}
\label{eq_tdepth1}
\end{equation}
the term I(b)/I(0) can be replaced by the quadratic limb-darkening profile \citep{mandelagol02} to give
\begin{equation}
D = \frac{R^{2}_{p}}{R^{2}_{\star}} \frac{(1- u_1 (1-\sqrt{1-b^2})-u_2 (1-\sqrt{1-b^2})^2)}{(1 - \frac{u_1}{3} - \frac{u_2}{6})}
\label{eq_tdepth2}
\end{equation}
\end{subequations}
where $u_1$ and $u_2$ are the quadratic limb-darkening coefficients \citep[same as $\gamma_1$ and $\gamma_2$ in][]{mandelagol02}. The limb-darkening parameters $v_1$ and $v_2$ in \mtq\ are linear combinations of the quadratic limb-darkening coefficients $u_1$ and $u_2$, with $v_1 = u_1+u_2$ and $v_2 = u_1-u_2$. These linear combinations were used while fitting the lightcurves, since it is known that directly fitting for limb-darkening coefficients results in strongly anti-correlated error distributions for various transit parameters \citep{brown01} and severely hinders the chance of Bayesian techniques from converging to accurate values (more in \sect\ \ref{sec_tmcmc}).

We wrote two versions of \mtq, one designed to fit transits observed with different filters (Multi-Filter), and the other designed to look for variations in the transit depth (Multi-Depth). If one makes transit observations using filters of different wavelength, the resulting set of transit lightcurves must be analyzed using a model that accounts for the different limb-darkening profiles, and differing transit depths extant in the data. The standard set of parameters used for Multi-Filter version of \mtq\ is $\boldsymbol{\theta}_{\text{Multi-Filter}}$ = \{$t_{T}$, $t_{G}$, $D_{j...N_F}$, $v_{1,j...N_F}$, $v_{2,j...N_F}$, $T_{i...N_T}$\}, where $T_{i}$ are the transit times and $v_1$ and $v_2$ are the limb-darkening parameters described in the previous paragraph. The subscripts ${i...N_T}$ and ${j...N_F}$ are used to denote multiple transits ($N_T$) and multiple filters ($N_F$) respectively. For \apostle's \tresb\ data we only observed using one filter, and the number of transit was eleven.

The second version of \mtq\ is designed to fit for the depths of each individual transit lightcurve separately. Variations in transit depth can arise for several reasons. Commonly invoked sources of such variations are starspots (cool photospheric regions) and faculae (hot photospheric regions) \citep[][and many more]{pont07,lanza09,knutson11}. The appearance and disappearance of spots on the stellar surface due to rotation or stellar activity cycles would result in variations in the transit depth. Transits occurring across the spotless stellar surface will be shallower than when active regions exist on the visible face of the star (for cool spots). Conversely, if these active regions are hot spots, the transit depths would change in the opposite manner. Other sources for transit depth variations include planetary oblateness, spin precession \citep{carterwinn10}, planetary rings \citep{barnesfortney04}, and satellites \citep{sartorettischneider99,tusnskivalio11}. The wide variety of proposed sources of transit depth variation means there may be several degeneracies to resolve if a transit depth variation is indeed detected. Nonetheless, understanding such phenomena may only be possible by establishing statistically significant measurements showing variable transit depths. The multi-filter capability of \mtq\ can be modified to fit for the depth of each transit as a unique parameter. In this case the set of parameters used is $\boldsymbol{\theta}_{\text{Multi-Depth}}$ = \{$t_{T}$, $t_{G}$, $D_{i...N_T}$, $v_{1,j...N_F}$, $v_{2,j...N_F}$, $T_{i...N_T}$\}, where $D_{i}$ is now fit for each transit instead of each filter. However, the filter corresponding to each depth is still tracked as the model needs to convolve the limb-darkening profile to correctly reproduce the full lightcurve profile. For \tresb\ observations only a single filter was used. In addition, one transit is chosen as the ``reference'' transit, using which \mtq\ internally computes several transit parameters such as $R_{p}/R_{\star}$, $a/R_{\star}$ etc. We picked transit \# 7 since it had high photometric precision, and was well-sampled.

In the sections that follow we describe results from our attempt to fit for transit parameters using both the Multi-Filter and Multi-Depth models. The following section outlines the Markov Chain Monte Carlo analyzer used in conjunction with \mtq.

\section{Transit MCMC} \label{sec_tmcmc}
Markov Chain Monte Carlo (MCMC) analyzers are now the standard for modeling data on exoplanets \citep{ford05,holman06,colliercameron07b,kundurthy11,gazak11}. We developed an MCMC routine called Transit MCMC (\tmcmc), which is designed to work in conjunction with \mtq. However, it can also be used to fit other models and data easily. The core \tmcmc\ routine uses the Metropolis-Hastings (M-H) algorithm, and is based on its implementation for astronomical data, as described by \citet{tegmark04} and \citet{ford05}. The Markov chain is computed by making jumps in parameter space and selecting those jumps which tend toward regions of parameter space at lower $\chi^2$. Jumps from the $j^{th}$ step in the chain are made with the following equation:
\begin{equation}
\boldtheta_{j+1} = \boldtheta_{j} +  \gaussian(0,\boldsymbol{\sigma}^2_{\theta}) f
\label{eq_jump}
\end{equation}
where $\boldtheta$ and $\boldsymbol{\sigma}_{\theta}$ are the vectors of model parameters and their associated step-sizes respectively. The term $\gaussian(0,\sigma_\theta^2)$, referred to as the proposal distribution, is a random number drawn from a normal distribution with a mean of 0 and a variance of $\sigma^2_{\theta}$. It is customary to use a Gaussian proposal distribution but not mandatory. We modify the M-H algorithm slightly by allowing adaptive jump-size adjustments that optimize the sampling rate of the Markov Chain integrator. The desired acceptance rates are achieved by adjusting the step-size controller ($f$) every 100 accepted steps according to $f_{\text{new}}$ = $W (f_{\text{old}}/N_{\text{trials}})$, where $N_{\text{trials}}$ are the number of steps attempted for the last 100 accepted steps and $W$ is a scaling factor which is 225 or 434 for single-parameter or multi-parameter chains respectively \citep[as noted in][]{colliercameron07b}. Given sufficient time, a converged chain will have traversed parameter space such that the ensemble of the points accurately represent the uncertainty distributions of the parameters in set $\boldtheta$.

\subsection{Markov Chains for \mtq} \label{sec_tmcmc4mtq}
For \apostle\ data sets, we explored system parameters using three different kinds of chains. Two of these were based on the Multi-Filter parameter set $\boldtheta_{\text{Multi-Filter}}$ described in \sect\ \ref{sec_mtq}; First, with Fixed Limb-Darkening Coefficients, and second, with Open Limb-Darkening Coefficients. The term limb-darkening coefficient will henceforth be abbreviated as \ldc. For the Fixed \ldc\ chains (FLDC), the coefficients were simply fixed to values tabulated for the appropriate observing filter \citep{claretbloemen11}. For the Open \ldc\ chains (OLDC), the limb-darkening parameters $v_{1}$ and $v_{2}$ are allowed to float. It has been shown that measured limb-darkening coefficients from high precision studies are in good agreement with tabulated values \citep{brown01,tingley06}. Another study using spectrophotometry from \hst\ STIS showed that any variations in the estimate of transit parameters due to inaccurate limb-darkening are lower than the 1-$\sigma$ uncertainty, with greater disagreement seen at shorter wavelengths \citep{knutson07}. Ground-based observations typically cannot achieve the level of photometric precision of space-based studies, hence constraining the limb-darkening from transit observations is difficult. Subtle inaccuracies in the fit limb-darkening profile can lead to incompatible estimates of system parameters, especially at short wavelengths \citep{kundurthy11}. In effect, the purpose of the Fixed \ldc\ and Open \ldc\ chains are simply to compare the differences in the fit limb-darkening coefficients to the tabulated values in the literature. The third type of Markov chain was run on the Multi-Depth parameter set $\boldtheta_{\text{Multi-Depth}}$ described in \sect\ \ref{sec_mtq}. \apostle\ lightcurves were gathered over a long time-baseline, and statistically significant depth variations seen in the data may help shed light on the various phenomena responsible for depth variations (see \sect\ \ref{sec_mtq}).

\begin{table}[!h]
\begin{center}
\caption{\label{table_bounds} Bounds applied for \mtqlong\ in \tmcmc}
\begin{tabular}{cl}
\hline \hline 
Bounds & Notes\\
\hline \hline
$t_T > 0$ & Non-zero transit duration \\
$t_G > 0$ & Non-zero limb-crossing duration \\
$D > 0$ & Non-zero transit depth \\
$ 1 - b^2 > 0$ & Impact parameters less than 1 (primary condition for transit) \\
$ b/(a/R_{\star}) \leq 1$ & Ensures real values for orbital inclination \\
$ 0 < u_1 < 1$ & Reasonable limb-darkening coefficients$^{*}$ \\
$ 0 < u_2 < 1$ & Reasonable limb-darkening coefficients$^{*}$ \\
\hline
\end{tabular}
\begin{tabular}{cl}
*applied when parameters were fit (OLDC chains) & \hspace{3.5in} \\
\end{tabular}
\end{center}
\end{table}

We applied bounds to several transit parameters and combinations of transit parameters in \mtq\ (as shown in Table \ref{table_bounds}). Most of these bounds were simply to check if the parameters had values that were physically realistic. For most parameters it was quite rare that the MCMC chains got close to the bounding limits, since the chains spend most of their time near low $\chi^2$ regions; best-fit parameter values for the \tres\ system are far from any physical bounds. When fitting for the limb-darkening however, the degeneracies proved very significant, and the Markov chains often strayed to unrealistic values. For example $u_1$ and $u_2$ took on values which would suggest limb-brightening rather than limb-darkening, either as a result of noisy data or due to gaps in the ingress or egress portions of the lightcurve. Limb-brightening is considered unphysical for broadband observations. Thus, after every jump (\eq\ \ref{eq_jump}), the vector of proposal parameters $\boldtheta_{j+1}$ is run through a series of sub-routines in \mtq, to check if any of the conditions in Table \ref{table_bounds} are violated. If any one of these conditions are violated, the vector is discarded and a new vector is generated. \tmcmc\ carries out this check until agreeable values emerge, and then continues with the rest of the algorithm.

\subsubsection{Executing and Analyzing Chains} \label{sec_postmcmc}
By varying the entire vector of model parameters, and applying a single step-size modifier (\eq\ \ref{eq_jump}), we run the risk of using mismatched step-sizes and hence undersampling the posterior distributions of some parameters. As mentioned before, well-constructed chains will properly sample posterior distributions given the correct acceptance rate \citep{gelman03}. The key to properly constructing a chain is to choose the relative starting step-sizes, for the parameter ensemble, such that they \textit{all} roam high and low probability regions of parameter space at roughly the same rate. Determining a reasonable set of starting step-sizes is done by running short exploratory Markov chains (40,000 steps) for each model parameter (holding all others fixed). If these exploratory chains have not stabilized to the optimal acceptance rate of $\sim 44\%$ \citep[as noted by][for single parameter chains]{gelman03} at the end of 40,000 steps, the chain is run until this rate is achieved. The jump sizes near the end of stabilized exploratory chains are then used as the starting steps for the multi-parameter chains. These multi-parameter chains will also be referred to as `long chains' from now on.

For each transiting system, we ran long chains of $2\times10^{6}$ steps from two different starting locations for each model scenario: Fixed \ldc, Open \ldc\ and Multi-Depth/Fixed \ldc. After completion we (1) cropped the initial stages of these chains to remove the burn-in phase, where the chain is far from the best-fit region, and (2) we exclude the stage where the chain is far from the optimal acceptance rate of 23$ \pm $ 5$\%$, as noted for multi-parameter chains \citep{gelman03}. We run three types of post-processing on the chains after cropping: (a) We compute the ranked and unranked correlations in the chains of every fit parameter with respect to the others. These statistics provide an estimate of the level of degeneracy between parameters in a given model. The next post-processing steps are two commonly used diagnostics to check for chain convergence, namely (b) computing the auto-correlation lengths and (c) the Gelman-Rubin \^{R}-static values \citep{gelmanrubin92}.

In a Markov chain, since a given step is only dependent on the preceding step, sections in the chain may show trends (\ie\ are correlated). The correlation length signifies the interval at which sampled points in a Markov chain will be uncorrelated. The effective length is the total number of points in the chain divided by the correlation length. Short correlation lengths (\ie\ large effective lengths) indicate that the MCMC ensemble represents a statistically significant and hence more precise sampling of the posterior distribution \citep{tegmark04}. Effective lengths $> 1000$ steps are commonly considered to be satisfactory. The \^{R}-statistic is computed using multiple chains of the same parameter set which have different initial conditions. An \^{R}-statistic close to 1 (to within 10$\%$) indicates that all chains have converged, cover approximately the same region of parameter space, and that the relevant parameter space has been sufficiently explored. Results from a chain are deemed useful if the auto-correlation and Gelman-Rubin conditions have been met.

\subsection{Comparing with the Transit Analysis Package}
It has been noted by \citet{carterwinn09} that transit lightcurves lacking any significant ``defects'' or artificial trends can have correlated noise buried within the overall scatter, which is invisible to visual examination. They test a wavelet based red-noise model on simulated transit lightcurves with and without artificial red-noise and find that models which do not fit for red-noise are subject to inaccuracies in transit parameters on the order of 2-3$\sigma$ and tend to have underestimated errors by up to 30$\%$. For transit timing studies, poor estimates such as these are cause for concern, since smaller errors and large deviations from the expected time can easily lead to false claims of TTVs. The detrending routine within \tmcmc\ removes long-term trends related to instrumental or other nuisance parameters. We visually examined the residuals of APOSTLE lightcurves (those observed during good conditions), and noted that the data may still have low-level correlated noise, even after detrending. The Transit Analysis Package \citep[TAP][]{gazak11} implements the red-noise model of \citet{carterwinn09}. We ran separate fits of transit parameters using TAP on the \textit{detrended} lightcurves from our \tmcmc\ fit.

The typical TAP parameter set is: $\boldsymbol{\theta}_{\text{TAP}}$ = \{$a/R_{\star}$, $i_{\text{orb}}$, $(R_{p}/R_{\star})_{i...N_F}$, $T_{i...N_T}$, $\sigma_{(\text{white},i...N_T)}$, $\sigma_{(\text{red},i...N_T)}$\}, where $\sigma_{(\text{white},i...N_T)}$ and $\sigma_{(\text{red},i...N_T)}$ are the white-noise and red-noise levels for $N_T$ transits, respectively. The TAP package does not fit for the period using the transit times, and often yields poor estimates of the period, so we fixed the period to that from \tmcmc\ fits. The limb-darkening was fixed to values from the literature. The orbital eccentricity and argument of periastron were kept fixed at 0 for \tresb.

One must also note a pitfall of the \citet{carterwinn09} red-noise model. Their fitting function expects the white-noise and red-noise components of a given lightcurve to be stationary, \ie\ there are no temporal variations in the Gaussian white-noise level and the red-noise's power-spectrum amplitude with time. They note that more elaborate noise models may be required to account for the fact that real data do not conform to these requirements. Typically the scatter is increased (due to lower signal-to-noise) at the portion of the lightcurve where the sky brightness is higher (like data taken close to twilight). Variable observing conditions may alter not just the white-noise properties of a lightcurve but also result in red-noise that is more complex than what can be described by \citet{carterwinn09}'s wavelet model. In addition, at the 1000 ppm level, stellar variability is not understood; it is reasonable to assume that the target and comparison may each contribute to the correlated noise in the lightcurve. In spite of these caveats, the TAP red-noise analysis serves as a good secondary check to results derived using \tmcmc\ (which does not account for red-noise).

\section{System Parameters}
\label{sec_sysparams}
As described in \sect\ \ref{sec_postmcmc} we ran three chains on lightcurves of \tresb: using the $\boldtheta_{\text{Multi-Filter}}$ parameter set with (1) Fixed \ldc, (2) Open \ldc, and (3) using the $\boldtheta_{\text{Multi-Depth}}$ parameter set with Fixed \ldc. The parameter sets are described in \sect\ \ref{sec_mtq}. All \tresb\ chains use $t_{G}$, $t_{T}$ and include the transit times $T_{i}$ for the 11 transits. Since all \tresb\ observations were taken in the \rband, the Fixed \ldc\ and Multi-Depth Fixed \ldc\ chains fit for 2 \ldc s for the \rband. The Multi-Depth chain was set up similarly, but with 11 free parameters for the transit depths. Post processing statistics and other data for these chains are listed in Table \ref{table_ChainStats_TRES3}. The column `$N_{\text{free}}$', `Chain Length', `Corr. Length' and `Eff. Length' list the number of free parameters, the length of the cropped chain, the correlation and effective lengths, respectively. All chains were run for approximately 2 million steps, but about 100,000 of the initial steps were removed to account for ``burn-in'' and selection rate stabilization. The Open \ldc\ and Multi-Depth chains both have quite low effective lengths indicating poor Markov chain statistics. Only the Fixed LDC, $\boldtheta_{\text{Multi-Filter}}$ model satisfies the condition of a well sampled posterior distribution (effective length is $> 1000$). The final two columns list the goodness of fit (\ie\ lowest $\chi^2$ in the MCMC ensemble) and Degrees-of-freedom (DOF) from the respective chain. Parameters from all chains had Gelman-Rubin \^{R}-statistics close to 1 indicating that the parameter space was covered evenly (though the OLDC and Multi-Depth FLDC chains were not sampled finely enough, based on the auto-correlation data).

\begin{table}[!h]\footnotesize
\begin{center}
\caption{\label{table_ChainStats_TRES3} \tmcmc\ Chains for TrES-3}
\begin{tabular}{cccccccc}
\hline \hline
Chain & Model Vector & N$_{\text{free}}$ &Chain Length & Corr. Length & Eff Length & $\chi^{2}$ & DOF \\
\hline
FLDC&$\boldtheta_{\text{Multi-Filter}}$&14&1,900,001&318&5,974&2503.67&2575\\
OLDC&$\boldtheta_{\text{Multi-Filter}}$&16&1,900,001&4,996&380&2519.14&2573\\
MDFLDC&$\boldtheta_{\text{Multi-Depth}}$&24&1,900,001&2,866&662&2692.94&2565\\
\hline
\end{tabular}
\end{center}
\end{table}

As noted previously in \citet{kundurthy11} the \rband\ limb-darkening is difficult to characterize using transit models. This result is seen again given the fact that the OLDC chain has the worst auto-correlation statistics. In addition the fact that \tresb\ has a near-grazing transit makes constraining the transit shape more challenging. The Multi-Depth models also did not converge, which could be due to the fact that several of the transits are not completely sampled during the eclipse event (see Figure \ref{figure_LC_TRES3}).

\begin{table}[!h]
\small
\begin{center}
\caption{\label{table_ParTable01_TRES3} TrES-3 Parameters for $\boldtheta_{\text{Multi-Filter}}$ }
\begin{tabular}{cccc}
\hline \hline
Parameter & FLDC & OLDC & Unit\\
\hline
\multicolumn{4}{c}{MTQ Parameters} \\
\hline
$t_{G}$&0.0210$\pm$0.0004&0.0212$^{+0.0007}_{-0.0006}$&days\\
$t_{T}$&0.0383$\pm$0.0002&0.0385$\pm$0.0010&days\\
$D_{\textrm{(r')}}$&0.0251$\pm$0.0002&0.0255$\pm$0.0003&-\\
$v_1{\textrm{(r')}}$&(0.6767)&0.7371$\pm$0.1279&-\\
$v_2{\textrm{(r')}}$&(0.3008)&-0.1782$\pm$0.4955&-\\
\hline
\multicolumn{4}{c}{Derived Parameters} \\
\hline
$(R_{p}/R_{\star})_{\textrm{(r')}}$&0.1652$\pm$0.0009&0.1649$\pm$0.0015&-\\
b&0.836$\pm$0.003&0.837$\pm$0.008&-\\
$a/R_{\star}$&5.97$\pm$0.03&5.91$^{+0.04}_{-0.05}$&-\\
$i_{orb}$&81.95$\pm$0.06&81.86$^{+0.08}_{-0.26}$&$^{o} (deg)$\\
$\nu/R_{\star}$&28.71$\pm$0.14&28.42$^{+0.19}_{-0.26}$&days$^{-1}$\\
$\rho_{\star}$&2.36$\pm$0.03&2.29$^{+0.05}_{-0.06}$&g/cc\\
P (1.3062 days +)&-1141$\pm$21&-1109$\pm$22&milli-sec\\
\hline
\end{tabular}
\end{center}
\end{table}
\begin{table}[!h]
\small
\begin{center}
\caption{\label{table_ParTable02_TRES3} TrES-3 Parameters for $\boldtheta_{\text{Multi-Depth}}$ }
\begin{tabular}{cccccc}
\hline \hline
Transit Depths & Value & Units & $R_{p}/R_{\star}$ & Value & Units \\
\hline
$(D)_{1}$&0.0249$\pm$0.0007&-&$(R_{p}/R_{\star})_{1}$&0.1676$\pm$0.0025&-\\
$(D)_{2}$&0.0261$\pm$0.0009&-&$(R_{p}/R_{\star})_{2}$&0.1711$\pm$0.0030&-\\
$(D)_{3}$&0.0276$\pm$0.0007&-&$(R_{p}/R_{\star})_{3}$&0.1754$\pm$0.0024&-\\
$(D)_{4}$&0.0256$\pm$0.0005&-&$(R_{p}/R_{\star})_{4}$&0.1696$\pm$0.0021&-\\
$(D)_{5}$&0.0237$\pm$0.0010&-&$(R_{p}/R_{\star})_{5}$&0.1638$\pm$0.0034&-\\
$(D)_{6}$&0.0261$\pm$0.0006&-&$(R_{p}/R_{\star})_{6}$&0.1710$\pm$0.0024&-\\
$(D)_{7}$&0.0259$\pm$0.0005&-&$(R_{p}/R_{\star})_{7}$&0.1705$\pm$0.0021&-\\
$(D)_{8}$&0.0269$\pm$0.0006&-&$(R_{p}/R_{\star})_{8}$&0.1734$\pm$0.0024&-\\
$(D)_{9}$&0.0274$\pm$0.0012&-&$(R_{p}/R_{\star})_{9}$&0.1750$\pm$0.0038&-\\
$(D)_{10}$&0.0254$\pm$0.0005&-&$(R_{p}/R_{\star})_{10}$&0.1690$\pm$0.0022&-\\
$(D)_{11}$&0.0277$\pm$0.0007&-&$(R_{p}/R_{\star})_{11}$&0.1758$\pm$0.0025&-\\
\hline
\multicolumn{6}{c}{Other MTQ Parameters} \\
\hline
Parameter & Value & Units & Parameter & Value & Units \\
\hline
$t_{G}$&0.0232$\pm$0.0007&days&$t_{T}$&0.1758$\pm$0.0025&days\\
$v_1{\textrm{(r')}}$&(0.6767)&-&$v_2{\textrm{(r')}}$&0.1758$\pm$0.0025&-\\
\hline
\multicolumn{6}{c}{Derived Parameters} \\
\hline
b&0.858$\pm$0.005&-&$a/R_{\star}$&5.96$\pm$0.03&-\\
$i_{orb}$&81.72$\pm$0.09&$^{o} (deg)$&$\nu/R_{\star}$&28.66$\pm$0.15&days$^{-1}$\\
$\rho_{\star}$&2.35$\pm$0.04&g/cc&-&-&-\\
\hline
\end{tabular}
\end{center}
\end{table}

\begin{figure}[!h] 
\begin{center} 
\includegraphics[width=0.95\textwidth]{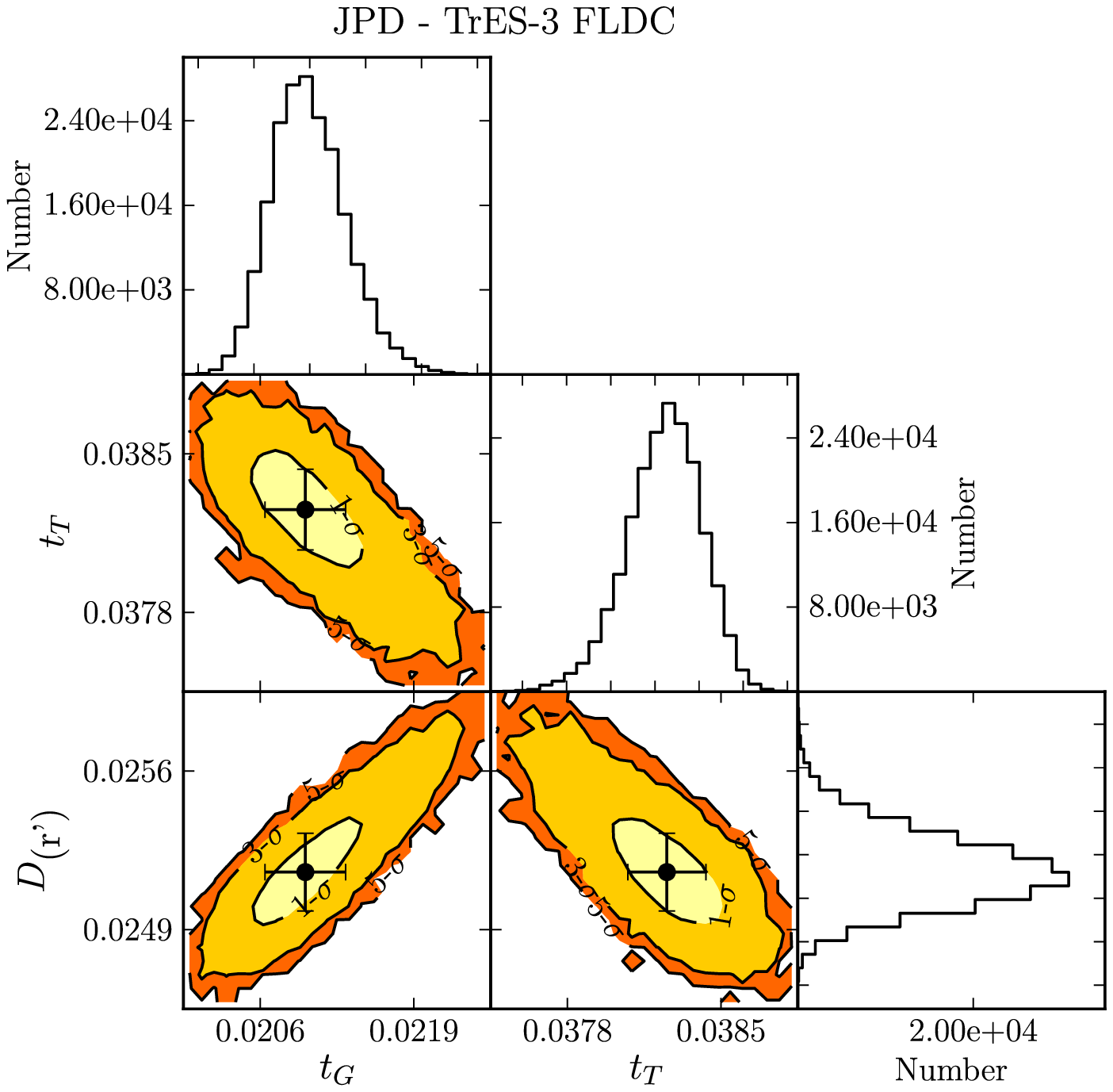}
\caption{\label{figure_MCMC_FLDC_TRES3}Plots of the Joint Probability Distributions (JPD) of parameters from the Fixed LDC chains, showing that due to the system's near-grazing transit the parameters chosen in $\boldtheta_{\text{Multi-Filter}}$ show correlations, unlike the cases for other systems discussed in this work. Table \ref{table_ParTable01_TRES3} gives units.}
\end{center} 
\end{figure}

\begin{figure}[!h] 
\begin{center} 
\includegraphics[width=0.95\textwidth]{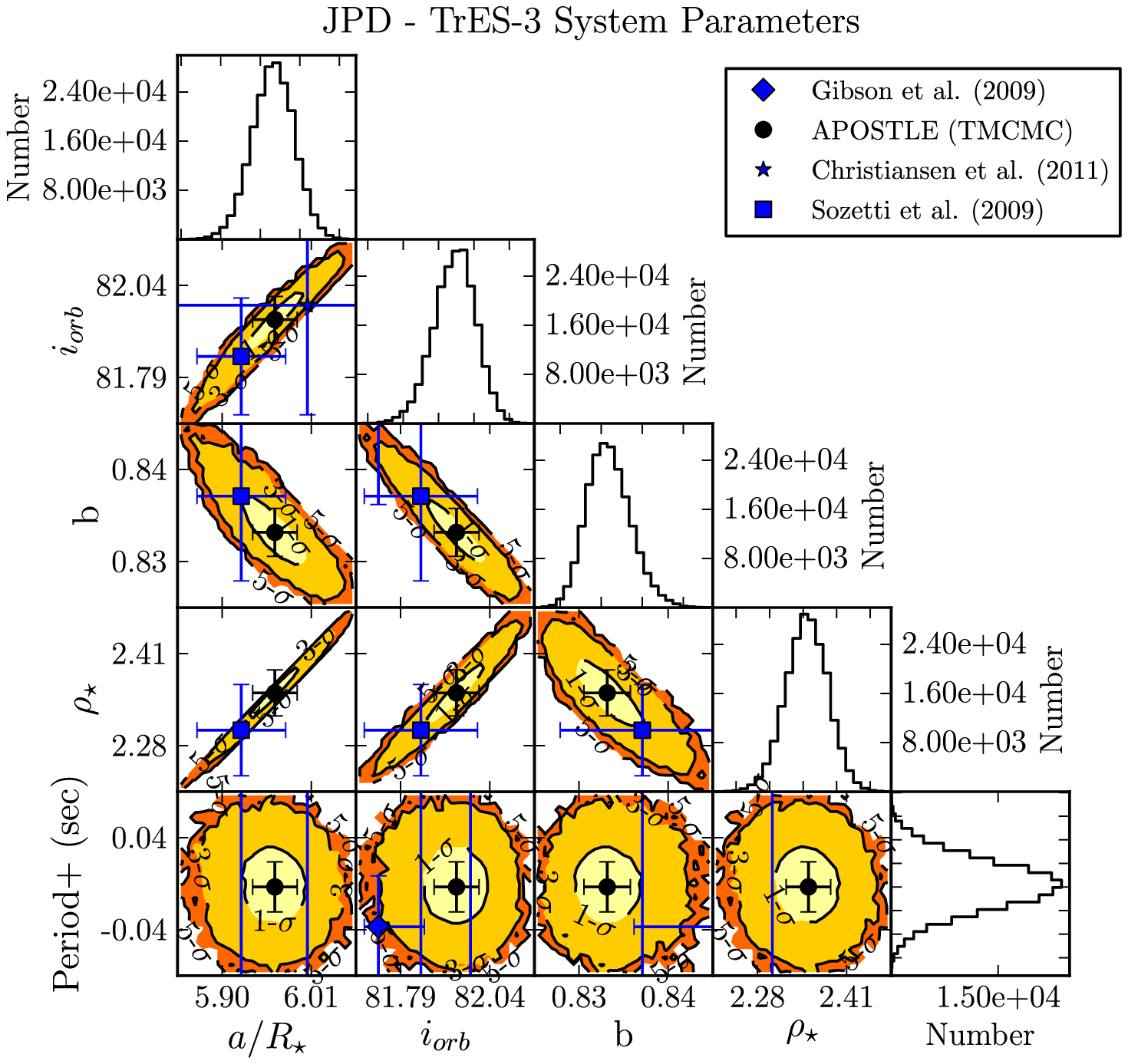}
\caption{\label{figure_MCMC_FLDC_DER_TRES3}Plots of the Joint Probability Distributions (JPD) of derived system parameters from the Fixed LDC chains. Parameter estimates available in the literature are overplotted. Table \ref{table_ParTable01_TRES3} gives units.}
\end{center} 
\end{figure}

The resulting best-fit parameter estimates are listed in Table \ref{table_ParTable01_TRES3} for the Multi-Filter models, and in Table \ref{table_ParTable02_TRES3} for the Multi-Depth models. These tables also list the derived system parameters. The transformation between the \mtq\ parameters to the derived system parameters are described in \citet{carter08} and \citet{kundurthy11}. Contour plots showing the joint probability distributions (JPDs) for the fit and derived parameters are shown in Figures \ref{figure_MCMC_FLDC_TRES3} and \ref{figure_MCMC_FLDC_DER_TRES3} respectively.

It is interesting to note that there is a clear degeneracy in the \mtq\ parameters as seen by the correlations in the posterior distributions of three parameters ($D$, $t_{G}$ and $t_{T}$) shown in Figure \ref{figure_MCMC_FLDC_TRES3}. From the impact parameter ($b$ = 0.84) we note, as previous studies have, that \tresb\ has a near-grazing transit. The correlation in the JPDs seen in Figure \ref{figure_MCMC_FLDC_TRES3} can be understood from Equation \ref{eq_tdepth2} and the following equation \citep{carter08}:
\begin{equation}
b = \sqrt{1 - (t_T/t_G)(R_{p}/R_{\star})}
\label{eq_b}
\end{equation}
Rearranging these two expressions gives an expression where $\frac{t_{G}}{t_{T}} \propto \frac{D}{1-b^2}$. \citet{carter08} have shown that when $b \rightarrow 1$ (\ie\ the transit is close to grazing), the fraction $t_{G}/t_{T}$ rises and the covariances between D, $t_{G}$ and $t_{T}$ rapidly deviate from zero. This degeneracy may be the primary reason why there is disagreement between the estimates of the system parameters for \tresb\ reported by various groups (see Table \ref{table_ParTable03_TRES3}). It may also explain the low statistical significance of the Open \ldc\ and Multi-Depth/Fixed \ldc\ chains, since Markov Chains take longer to converge when there are correlations between model parameters.

System parameters agree with previously published values in the literature, as seen by the overlap of the uncertainties in the JPD plot (Figure \ref{figure_MCMC_FLDC_DER_TRES3}). The errors from \tmcmc\ on the orbital inclination ($i_{\text{orb}}$), impact parameter ($b$) and stellar density ($\rho_{\star}$) are smaller than the previous best measurements of \citet{sozzetti09} by factors of 2.5, 3.5 and 2 respectively. However, since the \tmcmc+\mtq\ analysis does not include rednoise analysis the errors presented in Table \ref{table_ParTable02_TRES3} and Figure \ref{figure_MCMC_FLDC_DER_TRES3} are underestimates \citep{carterwinn09}. More conservative constraints were placed on a subset of these system parameters using the TAP package. Comparisons of some parameters and their uncertainties are presented in \ref{table_ParTable03_TRES3}.

\begin{table}[!h]
\small
\begin{center}
\caption{\label{table_ParTable03_TRES3} Comparison of Estimates of System Parameters for TrES-3b}
\begin{tabular}{ccccccc}
\hline \hline
Parameter & TMCMC & TAP & S09 & G09 & C11 & Units \\
\hline
$a/R_{\star}$ &5.97$\pm$0.03 & 5.89$\pm$0.05 & 5.93$\pm$0.06 & - & 6.01$\pm$0.84 & - \\
$i_{orb}$ &81.95$\pm$0.06 & 81.59$\pm$0.14 & 81.85$\pm$0.16 & 81.73$\pm$0.13 & 81.99$\pm$0.30 & $^{o} (deg)$ \\
b &0.836$\pm$0.003 & 0.861$\pm$0.007 & 0.840$\pm$0.010 & 0.852$\pm$0.013 & - & - \\
$\rho_{\star}$ &2.36$\pm$0.03 & 2.26$\pm$0.06 & 2.30$\pm$0.07 & - & - & g/cc \\
\hline
\end{tabular}
\footnotesize \begin{tabular}{l}
 TMCMC \& TAP values are from independent analysis of APOSTLE lightcurves \\
 S09 - \citep{sozzetti09}, G09 - \citep{gibson09}, C11 - \citep{christiansen11} \\
\end{tabular}\end{center}
\end{table}
It is clear that using TAP on the \apostle\ dataset and accounting for rednoise provides more conservative estimates of the system parameters. TAP errors are 10--30\% smaller than those reported by \citet{sozzetti09}. \citet{sozzetti09} and \citet{gibson09} account for correlated noise by scaling the photometric errors and generally have comparable uncertainties to the \apostle\ analysis with TAP. One must note that improved system parameters derived from transit measurements, like $\rho_{\star}$, can be used to place constraints on the age of the system \citep{sozzetti09,southworth10}, which in turn can be used to understand the evolutionary history of the exoplanetary atmosphere. As noted by \citet{carterwinn09}, we confirm that correlated noise in lightcurves is an obstacle for obtaining more precise estimates of system parameters.

\subsection{Transit Depth Analysis}
\label{sec_tdv}

The auto-correlation data indicate that the Multi-Depth/Fixed \ldc\ did not converge and hence errors for the parameters from this chain are unreliable. However, we can still study the results from the Multi-Depth fit assuming that the best-fit values are accurate. Since the Gelman-Rubin test indicate that the parameter space was fully traversed, the best-fit points (median values) from the chain are likely to be close to values that may have resulted from a converged chain.

\begin{figure}[!h] 
\begin{center} 
\includegraphics[width=0.95\textwidth]{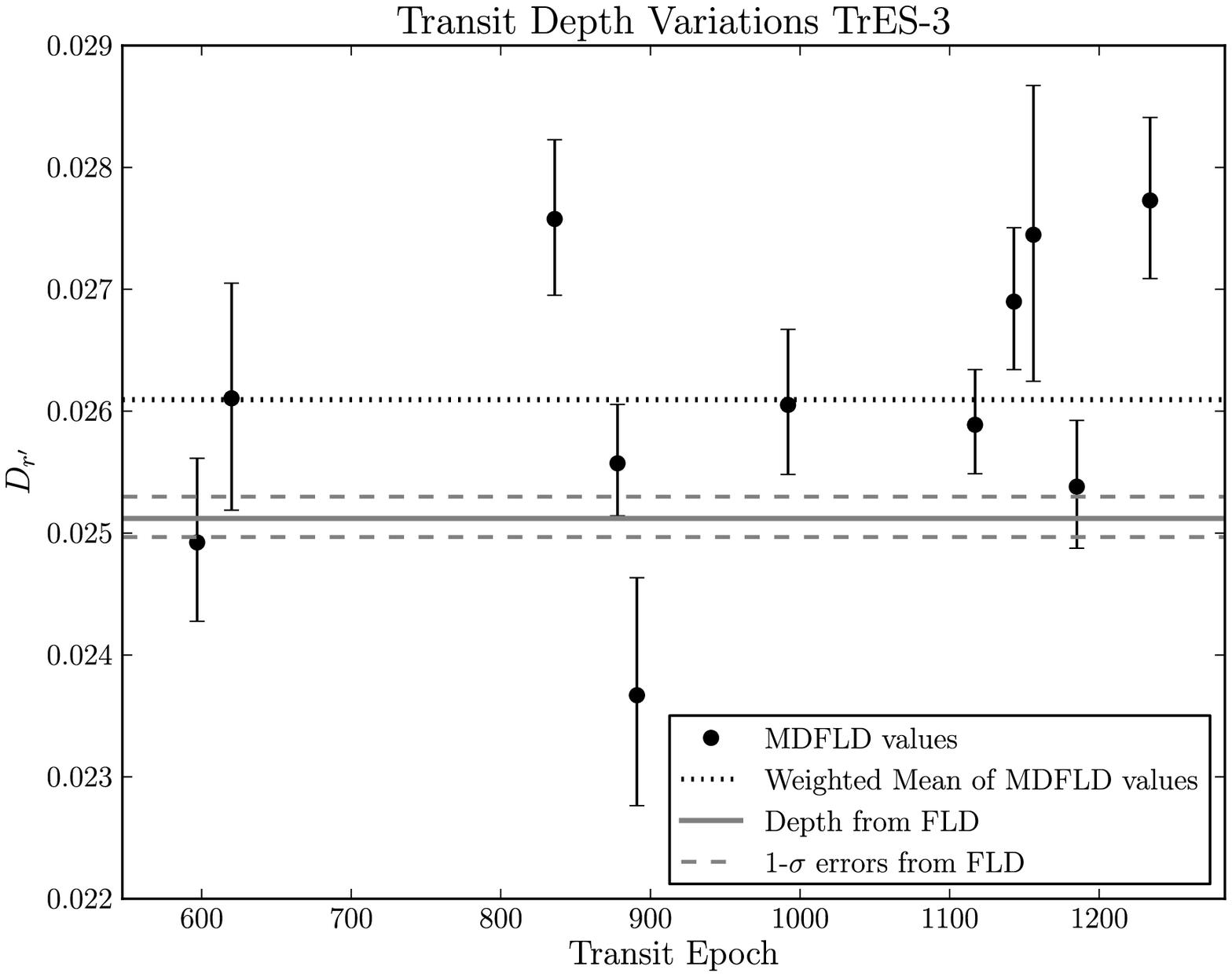}
\caption{\label{figure_TDepthV_TRES3}The transit depth $D$ as a function of transit number for \rband\ observations of \tresb. The solid horizontal and dashed lines represent the best-fit value and errors respectively for $D$ from the Fixed LDC \tmcmc\ fit. The dotted line is the weighted mean of transit depth values from the Multi-Depth Fixed LDC chains.}
\end{center} 
\end{figure}

Figure \ref{figure_TDepthV_TRES3} shows the transit depth vs. transit epoch for 11 \rband\ observations of \tresb. The overall variations in the \rband\ depth are 0.12\% compared to the 0.02\% uncertainty in D$_{(r')}$ from the joint fit to depths (FLDC, Table \ref{table_ParTable01_TRES3}). The median depth value (from depth measurements in Table \ref{table_ParTable02_TRES3} ) is 0.0261, compared to D$_{(r')} = $ 0.0251 from the FLDC chain. We note that several of the lightcurves are not completely sampled. We excluded lightcurve \# 3, 5, 8, 9 and 11 and recomputed the median depth and scatter to be 0.0257 and 0.04\% respectively, which is more consistent with the results from the Fixed \ldc\ chain. However, since the scatter is still larger than the uncertainty in the depth measured by the Single-Depth Fixed \ldc\ chain we cannot completely rule out variability. Since \tmcmc\ does not have rednoise analysis and the transit depth cannot be computed via TAP's parameter set we also cannot know the level at which the depth uncertainties in the Fixed LDC chains are underestimates. In addition, there are several facts about \tres\ which could support the notion that the depth variability seen in Figure \ref{figure_TDepthV_TRES3} is due to stellar activity. Firstly, the activity index reported for \tres\ classifies it as an active star \citep{sozzetti09,knutson10}. Secondly, observations in the \rband\ are known to be affected by spots, due to the fact that the $H\alpha$ line falls in this wavelength range, and spot-to-star contrast ratios are enhanced. However, there have been no indications of spot-crossing events in the transit data (a sure sign of stellar activity); this could be due to a large impact parameter and the greater likelihood of starspots being equatorial (assuming the stellar spin-axis is aligned with the sky-plane). Thirdly, the possibility that \tres\ is variable is evidenced by the changing flux-ratio with its comparison star (Column `Flux Norm.' in Table \ref{table_ObsSum_TRES3}); though this fact could just as easily be explained by variability in the comparison.

In summary, after accounting for incomplete and poor quality lightcurves, the resulting low-level variance in transit depth does not warrant a confident claim for the detection of variability in TrES-3. More continuous sampling of transit lightcurves, without interruptions would provide better insight into variability. Moreover, as with other parameter measurements, the affect of rednoise on depth measurements needs to be studied further.

\subsection{Transit Timing Analysis}
\label{sec_ttv}
Using Transit Timing Variations (TTVs) to look for additional planets was first proposed by \citet{agol05} and \citet{holmanmurray05}; in a system where only one planet is seen in transit, a deviation from the Keplerian period could indicate the presence of additional undetected planets. TTVs on the order of minutes can be produced if an unseen companion lies close to mean motion resonance with the transit planet \citep{holman10,lissauer11a,ballard11,nesvorny12}.

\begin{figure}[!h] 
\begin{center} 
\includegraphics[width=0.95\textwidth]{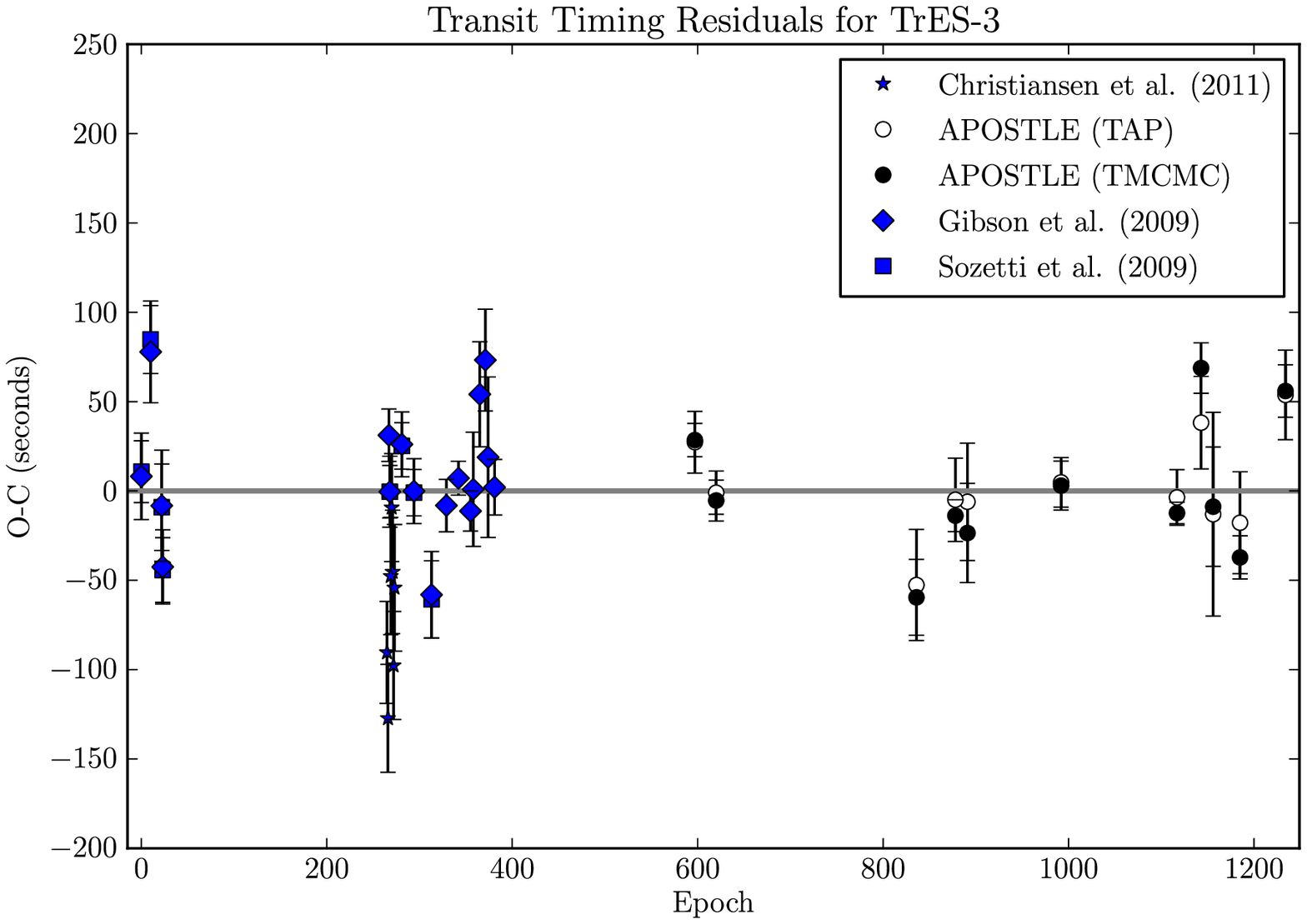}
\caption{\label{figure_OC_TRES3}The Observed minus Computed Transit Times for \tresb. Values from APOSTLE's \tmcmc\ fit, TAP and the literature are plotted. The horizontal axis represents the transit Epoch. The zero-line ephemeris is described in $\S$ \ref{sec_ttv}}
\end{center} 
\end{figure}

We gathered transit times published by \citet{sozzetti09}, \citet{gibson09} and \citet{christiansen11}, and pooled them alongside \tmcmc\ and TAP measurements of 11 \tres\ transit times. The timestamps of all \apostle\ data were converted to BJD (TDB) in the customized reduction pipeline \citep{kundurthy11}. The \apostle\ pipeline's time conversions have been verified by comparison to the commonly used time conversion routines made available by \citet{eastman10}. The transit times from the literature were also converted to BJD (TDB) before comparisons were made. The Observed minus Computed (O-C) plot is shown in Figure \ref{figure_OC_TRES3}. A linear ephemeris was fit to all the data using the equation,
\begin{equation}
T_i = T0 + \text{Epoch}_i \times P
\label{eq_transitephem}
\end{equation}
resulting in a best fit ephemeris of,
\begin{eqnarray*}
P = 1.306186483 \pm 0.000000070  & \text{days} \\
T0 = 2454185.9109932 \pm 0.0000502 & \text{BJD}
\end{eqnarray*}
with a goodness of fit $\chi^{2} =$ 204.1 for 52 Degrees-of-freedom. The large reduced chi-squared \redchi\ = 3.92 indicates that a linear fit does not precisely fit the transit times, and there are significant timing deviations in the data. The largest timing deviation of $\sim 127$ sec is from the \citet{christiansen11} data set. As previously noted, these deviations could either be due to underestimated timing errors or unaccounted timing systematics between different studies. The standard deviation of the O-C values is $\sim 43$ sec, hence the largest timing deviation is very close to being a 3-$\sigma$ outlier. Several features of the O-C plot seem to indicate that these timing variation are likely due to inconsistencies between the various methods used to derive transit times. For example, \citet{christiansen11}'s times are consistently early when compared to the expected transit times (\ie\ O-Cs are all negative). Removing the \citet{christiansen11} transit times from the ephemeris fit results in a millisecond change in the fit period, but improves the goodness of fit slightly to \redchi\ = 3.5, with the scatter being 35 seconds and the largest timing deviation being 78 sec, a $\sim$2-$\sigma$ offset. This fact establishes that there may be discrepancies between the methods used to derive the transit times, and hence we cannot claim any TTVs for \tres.

In order to compare the ephemerides derived with and without rednoise analysis, we fit for a linear ephemeris to the APOSTLE transit times from \tmcmc\ and TAP respectively (presented in the bottom half of Table \ref{table_OC_TRES3}). The difference between the periods derived for these subsets and the period derived from all available transit times was $< 12$ milli-seconds. The reduced $\chi^{2}$s were 7.99 and 1.33 for the \tmcmc\ and TAP subsets respectively, confirming that TAP gives more conservative errors for the transit times thanks to the red-noise analysis. Given the much more robust fit to a linear ephemeris from TAP's transit times we can rule out any TTVs in the system larger than $\sim 27.3$ sec (the scatter in the O-C for transit times derived from TAP). 

\begin{table}[!h]
\begin{center}
\caption{\label{table_OC_TRES3} APOSTLE Transit Times for TrES-3b}
\begin{tabular}{ccccc}
\hline \hline
Epoch & T0 (\tmcmc) & $\sigma_{T0}$ & T0 (TAP) & $\sigma_{T0}$ \\
 & 2,400,000+ (BJD) & (BJD) & 2,400,000+ (BJD) & (BJD) \\
\hline
597&54965.7046437&0.0001079&54965.7046300&0.0002000\\ 
620&54995.7465390&0.0001326&54995.7465900&0.0001400\\ 
836&55277.8821699&0.0002456&55277.8822500&0.0003600\\ 
878&55332.7425269&0.0001033&55332.7426300&0.0002700\\ 
891&55349.7228375&0.0003216&55349.7230400&0.0003800\\ 
992&55481.6479691&0.0001589&55481.6479900&0.0001600\\ 
1117&55644.9210892&0.0000699&55644.9211900&0.0001800\\ 
1143&55678.8828740&0.0001637&55678.8825200&0.0003000\\ 
1156&55695.8623990&0.0003861&55695.8623500&0.0006600\\ 
1185&55733.7414753&0.0001397&55733.7417000&0.0003300\\ 
1234&55797.7456860&0.0001699&55797.7456600&0.0002900\\ 

\hline 
\end{tabular}
\begin{tabular}{ccccc}
Fit & Period (days) & $\sigma_{\text{P}}$ & T0 (BJD) & $\sigma_{\text{T0}}$ \\
 \hline
\tmcmc & 1.306186240 & $\pm$ 0.000000181 & 2454965.7043612 & $\pm$ 0.0000770 \\
TAP & 1.306186489 & $\pm$ 0.000000302 & 2454965.7043416 & $\pm$ 0.0001118 \\
\hline
\end{tabular}
\end{center}
\end{table}

For the case when planets are in mean-motion resonance (MMR), \citet{agol05} show that the analytic expression, $\delta t_{\text{max}} \sim \frac{P}{4.5 j} \frac{m_{\text{pert}}}{(m_{\text{pert}}+m_{\text{trans}})}$, can roughly estimate the amplitude of the timing deviation ($\delta t_{\text{max}}$). The quantities $m_{\text{pert}}$, $m_{\text{trans}}$, $P$ and $j$ are the mass of the unseen perturber, the mass of the transiting planet, the orbital period of the transiting planet and the order of the resonance respectively. For the \tres\ system, we can rule out possible system configurations given the variations seen in the O-C result from \apostle's TAP fit. The maximum possible TTV amplitude that could be hidden within the variations seen in \apostle\ transit times are $\sigma_{TT,TAP} \sim \delta t_{\text{max}} \sim 27 \text{sec}$. Using the orbital period from Table \ref{table_OC_TRES3} and the mass of \tresb, $m_{\text{trans}}$ = 1.91 $M_{\text{Jup}}$ \citep{sozzetti09}, we compute the maximum mass perturber that could exist in the \tres\ system in the 2:1 MMR to be $\sim 0.66 M_{\earth}$, \ie\ additional planets with $M_{p} < 0.66$ \mearth may exist near the 2:1 MMR. At higher order resonances, this maximum mass (for a possible perturber) is larger. 

\section{Conclusions}
\label{sec_conclusions}

\vspace{\baselineskip} \N \textbf{Photometric Precision:} \apostle\ monitored \tresb\ over a period of two years between 2009 and 2011, gathering 11 \rband\ transit lightcurves. \apostle\ achieved photometric precision between 600--1200ppm (excluding nights with poor seeing). The summary of observational results is presented in Table \ref{table_ObsSum_TRES3}.

\vspace{\baselineskip}  \N \textbf{\tresb\ System Parameters:} From our analysis of 11 lightcurves of \tresb, we were able to confirm previous estimates of system parameters for the \tres\ system. Our estimates of derived system parameters in Table \ref{table_ParTable01_TRES3} show improvements ($\sim$ 10--30\%) from previous measurements. Due to the system's near grazing transit we note that several free parameters from \mtq\ showed strong correlations in their Joint-Probabilities (see Figure \ref{figure_MCMC_FLDC_TRES3}). Such correlations are not suited for rapid MCMC convergence, yet we were able to produce Markov chains that converged and hence are able to derive statistically significant uncertainties for system parameters; see results from the FLDC chain in Table \ref{table_ParTable01_TRES3}. As previous attempts showed \citep{kundurthy11}, the Markov chains where limb-darkening coefficients were set as free parameters failed to converge, and so we report that \apostle\ did not achieve the photometric precision required to constrain limb-darkening coefficients for \tres.

\vspace{\baselineskip}  \N \textbf{Search for Transit Depth Variations:}
Variations in transit depth over epoch could be evidence for stellar variability. \tres\ is known to be variable \citep{sozzetti09}, hence one might have expected strong variations in the transit depth with epoch. Our uncalibrated flux ratios between the target and comparison show significant variability, yet we cannot rule out variability in the comparison star as the cause. Our Multi-Depth fits also show variations in the transit depth over transit epoch (see Figure \ref{figure_TDepthV_TRES3} and Table \ref{table_ParTable02_TRES3}), however we refrain from making a confident assertion on the detection of stellar variability since, (a) the parameters from the Multi-Depth chain do not have statistically significant errors as the chain failed to converge, and (b) we cannot rule out inaccuracies due to the incomplete sampling of several transits, and (c) the near-grazing nature of this transit makes constraining the depth of the transit trough more challenging. In fact removing several of the incompletely sampled transits lowers the overall scatter in transit depth measurements.

\vspace{\baselineskip}  \N \textbf{Search for Transit Timing Variations:} The transit timing precisions achieved by \apostle\ easily allow for the detection of TTV signals $>$ 1min (see TTV uncertainties in Table \ref{table_OC_TRES3}). This is a direct consequence of the ability of the \apostle\ program to gather lightcurves with high photometric precision from the ground. However, we were unable to detect significant timing variations for \tresb\ in our data. We find that fitting for transit parameters while accounting for red-noise provided more conservative error estimates on transit times. Transit times derived using \tmcmc\ showed significant deviations from a linear ephemeris fit. The same data analyzed with analyzed with TAP \citep{carterwinn09,gazak11} significantly reduced these deviations. The overall scatter in the O-C values from our rednoise analysis was on the order of $\sim 27$ sec, which rules out planetary companions more massive than 0.66 \mearth near the 2:1 MMR, and larger companions near higher order resonances.

Transit times published in the literature are derived using different techniques for fitting transit parameters. A proper analysis of transit times would need a simultaneous analysis of transit lightcurves using a transit model that is (1) suited for Bayesian inference \citep[\ie\ with a fairly uncorrelated parameter set,][]{carter08} and (2) a transit model that can adequately account for red-noise in the data \citep[like TAP,][]{gazak11}. The catalog of transit times accumulated in the literature and used in this study of \tresb\ are not derived from such an analysis. MCMC analysis can also be inefficiently slow given large parameter sets and complex internal checks on a given model. Fast MCMC routines like The MCMC Hammer \citep{foreman-mackey12} may in the future, allow for rapid analysis using complex models. \textit{We provide the entire set of \apostle\ lightcurves of \tresb\ with this publication to allow future projects to apply improved methods to study this system.} 

Extrapolations from \keptel\ planetary candidate data seem to indicate that small planets may be ubiquitous \citep{borucki11b}. Interesting trends that have been noted are, (1) Hot-Jupiters tend be alone, \ie\ lacking other transiting planet siblings \citep{latham11,steffen12} and (2) members of multi-planet systems with short period planets (Period $<$ 10 days) are more likely to be Hot-Neptunes \citep{latham11,lissauer11b}. This trend seems to reveal the existence of different formation pathways among volatile-rich planets. Hence the lack of detections TTVs in \tresb\ (a Hot-Jupiter) is consistent with \keptel's findings.

\section*{Acknowledgments}
Funding for this work came from NASA Origins grant NNX09AB32G and NSF Career grant 0645416. RB acknowledges funding from NASA Astrobiology Institute's Virtual Planetary Laboratory lead team, supported by NASA under cooperative agreement No. NNH05ZDA001C. Data presented in this work are based on observations obtained with the Apache Point Observatory 3.5-meter telescope, which is owned and operated by the Astrophysical Research Consortium. We would like to thank the APO Staff, APO Engineers, Anjum Mukadam and Russel Owen for helping the APOSTLE program with its observations and instrument characterization. We would like thank S. L. Hawley for scheduling our observations on APO. This work acknowledges the use of parts of J. Eastman's EXOFAST transit code as part of APOSTLE's transit model \mtq. We would also like to thank an anonymous referee for providing feedback that helped improve our paper.

\end{document}